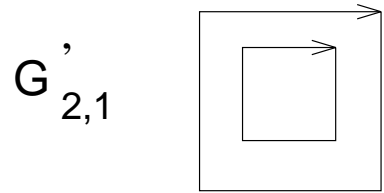

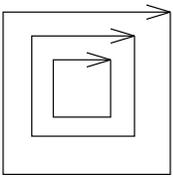 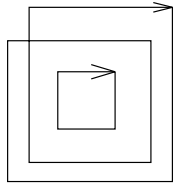 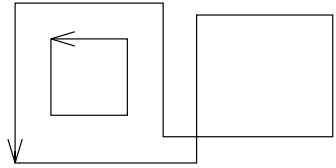 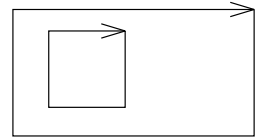

Fig. 1

# QCD$_3$ vacuum wave function

Qi-Zhou Chen,[1,2] Xiang-Qian Luo,[3,4,5] and Shuo-Hong Guo[2]

[1]*China Center of Advanced Science and Technology (World Laboratory),
P.O. Box 8730, Beijing 100080, People's Republic of China*

[2]*Department of Physics, Zhongshan University, Guangzhou 510275, China*

[3]*Departamento de Física Teórica, Facultad de Ciencias,
Universidad de Zaragoza, 50009 Zaragoza, Spain*

[4]*HLRZ, Forschungszentrum, D-52425 Jülich, Germany* \*

[5] *Deutsches Elektronen-Synchrotron DESY, D-22603 Hamburg, Germany*

**Abstract**

We investigate quantum chromodynamics in 2+1 dimensions (QCD$_3$) using the Hamiltonian lattice field theory approach. The long wavelength structure of the ground state, which is closely related to the confinement phenomenon, is analyzed and its vacuum wave function is evaluated by means of the recently developed truncated eigenvalue equation method. The third order estimations show nice scaling for the physical quantities.

---

\*Mailing Address

QCD has been accepted to be the most successful gauge theory of strongly interacting particles. In QCD, hadronic matter is composed of quarks, and interactions between them are mediated by eight massless gluons generated by the SU(3) gauge group. Asymptotic freedom of QCD at short distances makes the perturbative calculations of high energy processes possible. At long distances, however, there are a lot of low energy fundamental properties like confinement of quarks and gluons, vacuum structure, chiral-symmetry breaking, glueball masses, hadronic spectrum and weak interaction processes, and behaviors of hadronic matters at high temperature or high density, which can not be studied perturbatively.

Over the past two decades, lattice gauge theory (LGT) has developed into a promising first principle nonperturbative approach to these phenomena. QCD in the pure gauge sector possesses a nontrivial vacuum structure and bound states called glueballs. The Hamiltonian LGT provides a convenient tool for the estimations of the wave functions for the ground state and excited states. Furthermore, physical observables such as glueball masses correspond to the eigenvalues. In this context, both numerical [1, 2, 3] and analytical [4, 5, 6, 7, 8, 9] efforts have been made. In [3], Arisue argues that the wave function of the ground state of a D-dimensional non-abelian theory for the long wavelength configurations $A$ is

$$\Psi(A) = exp[-\frac{\mu_0}{e^2} \int d^{D-1}x \ tr F^2 - \frac{\mu_2}{e^6} \int d^{D-1}x \ tr(\mathcal{D}F)^2], \quad (1)$$

with $e$ being the renormalized coupling, $F$ the field strength tensor and $\mathcal{D}$ the covariant derivative. The idea was confirmed by his Monte Carlo simulation of a (2+1)-dimensional SU(2) lattice model. As well explained in [3], the correlation length of the continuum gauge field strength tensor in the vacuum, given by the square root of the ratio of the coefficients in (1), has the order of $1/e^2$, the same order of the confinement scale of the theory.

The motivations for the investigations of the (2+1)-dimensional lattice models are as follows.
(1) they have potential applications to high temperature superconductivity;
(2) they have many similarities to QCD in 3+1 dimensions (QCD$_4$) like asymptotic freedom, quark confinement, spontaneous chiral-symmetry breaking, and meson and glueball spectrum. It would be more economical to test various techniques on such a nontrivial theory which has as much as the same properties as QCD$_4$ but simpler (due to the advantage of lower dimensionality and superrenormalizability).

Recently, we proposed an analytical method [6, 7] for understanding the long wavelength behavior of Hamiltonian LGT, which is similar to Greensite's method [4] of truncated eigenvalue equation but with a different truncation scheme. The vacuum wave function of (2+1)-dimensional SU(2) gauge theory were evaluated up to the third and fourth orders, and nice scaling behavior and agreement with the Monte Carlo data [3] were observed.

It is very desirable to extend the method to a more realistic gauge group: SU(3). As will be seen, because of the nature of the group, the classification of the graphs is more complicated. To our knowledge, there has not been a published work on detailed investigation of the long wavelength vacuum structure of pure $QCD_3$ at zero temperature. There exist only preliminary analytic calculations in $QCD_3$ with fermions [10], in addition to some analytic analysis in the continuum [11] and Monte Carlo data [12, 13] for quenched $QCD_3$ at finite temperature. Recently, there has been an attempt [14] to include dynamical fermions in the numerical simulation of $QCD_3$.

The purpose of this paper is to describe and further explore our method for studying the vacuum structure of $QCD_3$ in the pure gauge sector. The study of the ground state properties is a first step towards the understanding of the structure of the glueballs and hadrons. The starting point is the discretization of the Yang-Mills theory in the Hamiltonian formulation (discrete in space, continuous in time, and temporal gauge $A(x, k_t) = 0$)

$$H = \frac{g^2}{2a} \sum_l E_l^\alpha E_l^\alpha - \frac{1}{ag^2} \sum_p Tr(U_p + U_p^\dagger - 2), \qquad (2)$$

where $g$ is the bare coupling, $a$ is the lattice spacing, $E_l^\alpha = E^\alpha(x, k)$ is the color-electric field on the link $l$ at site $x$ and positive direction $k$, and the second term is the color-magnetic energy with $U_p$ being the product of link variables $U_l = exp[iga A(x, k)]$ around an elementary plaquette. In 2+1 dimensions, the bare coupling and the lattice spacing have a simple relation $g^2 = e^2 a$. This Hamiltonian can be derived directly from Wilson's lattice Lagrangian using either the transfer matrix or canonical transformation. The gauge fields have to satisfy the commutation relations

$$[E^\alpha(x, k), U(y, j)] = T^\alpha U(x, k) \delta_{x,y} \delta_{k,j},$$

$$[E^\alpha(x,k), U^\dagger(y,j)] = -U^\dagger(x,k)T^\alpha \delta_{x,y}\delta_{k,j}, \tag{3}$$

with $T^\alpha$ being the fundamental representation of the $\alpha$th generator of the gauge group.

The wave function of the ground state is assumed to be of the form

$$|\Omega\rangle = exp[R(U)]|0\rangle, \tag{4}$$

where the bare vacuum $|0\rangle$ is defined to be fluxless and $R(U)$ consists of gauge invariant operators such as the Wilson loops. The vacuum state with energy $\epsilon_\Omega$ has to satisfy the lattice Schrödinger equation

$$H|\Omega\rangle = \epsilon_\Omega|\Omega\rangle, \tag{5}$$

which results in an eigenvalue equation for $H$

$$\sum_l \{[E_l,[E_l,R(U)]] + [E_l,R(U)][E_l,R(U)]\} - \frac{2}{g^4}\sum_p Tr(U_p + U_p^\dagger) = \frac{2a}{g^2}\epsilon_\Omega. \tag{6}$$

This equation can be solved by a truncation method [6, 7], in which $R(U)$ is expanded in order of graphs,

$$R(U) = \sum_i R_i(U), \tag{7}$$

and the order is defined as the number of plaquettes involved. Denote $R_i$ and $R_j$ as the graphs of order $i$ and $j$ respectively, and all new graphs created by $\sum_l [E_l,R_i(U)][E_l,R_j(U)]$ are defined as graphs of order $i+j$. Then the $n$th order truncated eigenvalue equation is

$$\sum_l \{[E_l,[E_l,\sum_i^n R_i(U)]] + \sum_{i+j\leq n}[E_l,R_i(U)][E_l,R_j(U)]\} - \frac{2}{g^4}\sum_p Tr(U_p + U_p^\dagger)$$

$$= \frac{2a}{g^2}\epsilon_\Omega. \tag{8}$$

The long wavelength limit of a graph is obtained by small $a$ expansion of the graph. In this limit, the vacuum wave function (4) is reduced to the continuum one (1). The ground state of the form (4) implies that at large scales, the pure gauge lattice vacuum is governed by multi monopole configurations.

In SU(2) gauge group, $TrU_p = TrU_p^\dagger$ and all loops with crossing can be transformed into loops without crossing. Then according to this rule, there are one graph of first order, three graphs of second order, and nine graphs of third order. In SU(3), however, these are no longer the case.

In the analytical calculation, the unitary and unimodular conditions lead to constrains on the graphs. Any group element $A$ of SU(3) has to satisfy the following condition

$$A_{i_1 j_1} A_{i_2 j_2} A_{i_3 j_3} \epsilon_{j_1 j_2 j_3} = \epsilon_{i_1 i_2 i_3}, \tag{9}$$

where a summation over the repeated indices is implied. We rewrite this condition as

$$2(A^\dagger)_{ij} = 2(A^2)_{ij} - 2A_{ij} TrA + [(TrA)^2 - Tr(A^2)]\delta_{ij}, \tag{10}$$

or

$$2\delta_{ij} = 2(A^3)_{ij} - 2(A^2)_{ij} TrA + [(TrA)^2 - Tr(A^2)]A_{ij}, \tag{11}$$

from which the relations among different graphs can be established. For example,

$$2G_1^\dagger = G'_{2,1} - G_{2,1},$$

$$6 = G'_{3,1} - 3G'_{3,2} + 2G_{3,1}$$

$$G_{2,4} = G_{2,3} - G_{3,6} + G'_{3,3}$$

$$G_{2,5} = G_{2,6} - G_{3,3} + G'_{3,4},$$

$$..., \tag{12}$$

where $G'$ and $G$ are the graphs which can be found in Fig. 1 and Fig. 2 with $G_1 = \Box = TrU_p$ and so on. One sees that not only graphs of the same order, but also graphs of different orders mix, so that the classification becomes rather involved. In this paper, we choose as much as possible the connected graphs (e.g. $G_{21}$) as independent graphs, and try our best to transform the disconnected ones (e.g. $G'_{2,1}$) into the connected ones, because we think that the connected ones give more relevant physical information at a given order. The complete set of graphs up to the third order are

$$R_1(U) = C_1 G_1 + h.c.,$$

$$R_2(U) = \sum_{i=1}^{6} C_{2,i} G_{2,i} + h.c.,$$

$$R_3(U) = \sum_{i=1}^{29} C_{3,i} G_{3,i} + h.c., \tag{13}$$

which graphs are plotted in Figs. 2.1, 2.2 and 2.3 respectively. Substituting them into (8), we obtain the 36 nonlinear equations for the coefficients $C_1$, $C_{2,i}$ and $C_{3,i}$.

Physical quantities like $\mu_0$ and $\mu_2$ are related to the coefficients of the graphs in the long wavelength limit:

$$\mu_0 = [C_1 + 6(C_{2,2} + C_{2,4} + C_{2,5}) + 4(C_{2,1} + C_{2,3}) + 15(C_{3,2} + C_{3,4} + C_{3,5})$$

$$+ 27(C_{3,7} + C_{3,12} + C_{3,13} + C_{3,14} + C_{3,19} + C_{3,20} + C_{3,29})$$

$$+ 3(C_{3,15} + C_{3,16} + C_{3,24} + C_{3,25} + C_{3,28}) + C_{3,6} + C_{3,17} + C_{3,18} + C_{3,23} + C_{3,26}$$

$$+ 15(C_{3,8} + C_{3,10} + C_{3,11} + C_{3,21} + C_{3,22}) + 9(C_{3,9} + C_{3,3} + C_{3,1} + C_{3,27})]g^4,$$

$$\mu_2 = [C_{3,6} - C_{3,3} + \frac{3}{2}(C_{3,15} + C_{3,16} + C_{3,24} + C_{3,25} + C_{3,28})$$

$$-\frac{3}{2}(C_{3,8} + C_{3,10} + C_{3,11} + C_{3,21} + C_{3,22})$$

$$-3C_{3,9} - C_{3,17} + 2C_{3,18} + C_{3,23} - 2C_{3,27} + \frac{1}{2}(C_{2,6} - C_{2,3})]g^8. \qquad (14)$$

$\mu_0$ and $\mu_2$ should be constants in the weak coupling limit $g \to 0$ or $\beta = 6/g^2 \to \infty$ as required by the renormalizability of the theory. Figure 3 gives a comparison between the third order results from the strong coupling expansion and the truncated eigenvalue equation. (The solid line is made by joining 100 data in the interval $\beta \in [0.12, 12]$, while the crosses are only the representative points.) They are consistent in the strong coupling region ($\beta < 6$), which implies that the calculation using the truncated eigenvalue equation method is supported by the results from the strong coupling expansion. For larger $\beta$, it is not surprising that the strong coupling expansion method no longer works. It is usually hoped that beyond the strong coupling region, there is a scaling region for extracting continuum information when the physical quantities become approximately constants. From the intermediate coupling ($\beta \approx 6.84$) till the weak coupling ($\beta \approx 11.52$), the data from the the truncated eigenvalue equation method show nice scaling behavior, thus suggesting the correct long wavelength continuum limit (1) of the vacuum wave function (4). From the results we estimate

$$\mu_0 \approx 0.5411 \pm 0.0038,$$

$$\mu_2 \approx -0.0781 \pm 0.0024, \qquad (15)$$

where the mean values are the averaged ones over the 40 data in the scaling region $\beta \in [6.84, 11.52]$, while the error analysis is based on the jackknife method (only a rough evaluation for the errors). As far as we know, the data for $\mu_0$ and $\mu_2$ from Monte Carlo simulations or other analytic methods are still lacking.

The presence of the expected scaling behavior in the intermediate and weak coupling regions tells us that even at finite bare coupling $g$, it is possible to extract the continuum information along the line of constant physics, since it will eventually flow into the critical point $g_{cr} = 0$, i.e., the continuum limit when the cut-off becomes larger and larger. These results again indicate that the correlation length of the continuum gauge field strength tensor in the ground state of $QCD_3$ has the order of $1/e^2$, i.e., the order of the confinement scale. Of course, as the continuum limit $a \to 0$ or equivalently $\beta \to \infty$ is approached, the correlation length in the lattice unit will be divergent so that the inclusion of higher and higher orders of the graphs is required to better represent the vacuum state.

There remains the problem about the choice of set of independent graphs. Due to the fact that elements of the gauge group have to satisfy the unitary and unimodular conditions, graphs of different orders mix. There is ambiguity in choosing a set of independent graphs at a given order. Different choice of independent graphs might give different results when truncating at a finite order. In this paper, we have chosen as much as possible the connected graphs, because we think that these graphs may stand for more coherence, and may lead to more rapid convergence to the continuum limit. Mixing with lower order graphs also appeared in [5] as a result of the shifting procedure. Here we invoke the unimodular condition for the necessity of mixing. We think that mixing is essential and there might exist some certain scheme for most efficient approach to scaling. This problem is currently under close investigation.

In summary, we have successfully applied the truncated eigenvalue equation method [6] to the realistic group SU(3) with some new prescription scheme for the classification of the graphs, and determined nonperturbatively the ground state wave function. The observation of scaling for $\mu_0$ and $\mu_2$ indicates the correct continuum behavior of the lattice vacuum wave function at relatively large scale or long wavelength. Once a reasonable form of the ground state is established, other physical quantities such as the spectrum of the excited states can be evaluated. Extension of our $QCD_3$ model and techniques to 3+1 dimensions is hopeful to yield phenomenologically relevant results. Such work is in progress [15].

Q.Z.C. and S.H.G. were supported by the Doctoral Program Foundation of Institute of Higher Education, the People's Republic of China and the Advanced Research Center of Zhongshan University, Hong Kong. X.Q.L. thanks DESY, Germany for support.

**Figure Captions**

Fig. 1. Graphs $G'$ in (12).

Fig. 2.1. First order graph in $R(U)$.

Fig. 2.2. Second order graphs in $R(U)$.

Fig. 2.3(a). Third order graphs in $R(U)$.

Fig. 2.3(b). Third order graphs (continuation of Fig. 2.3(a)) in $R(U)$.

Fig. 3. $\mu_0$ and $\mu_2$ as a function of $\beta$, where the triangles stand for the results from the strong coupling expansion, while the crosses are those from the third order calculation of the truncated eigenvalue equation method.

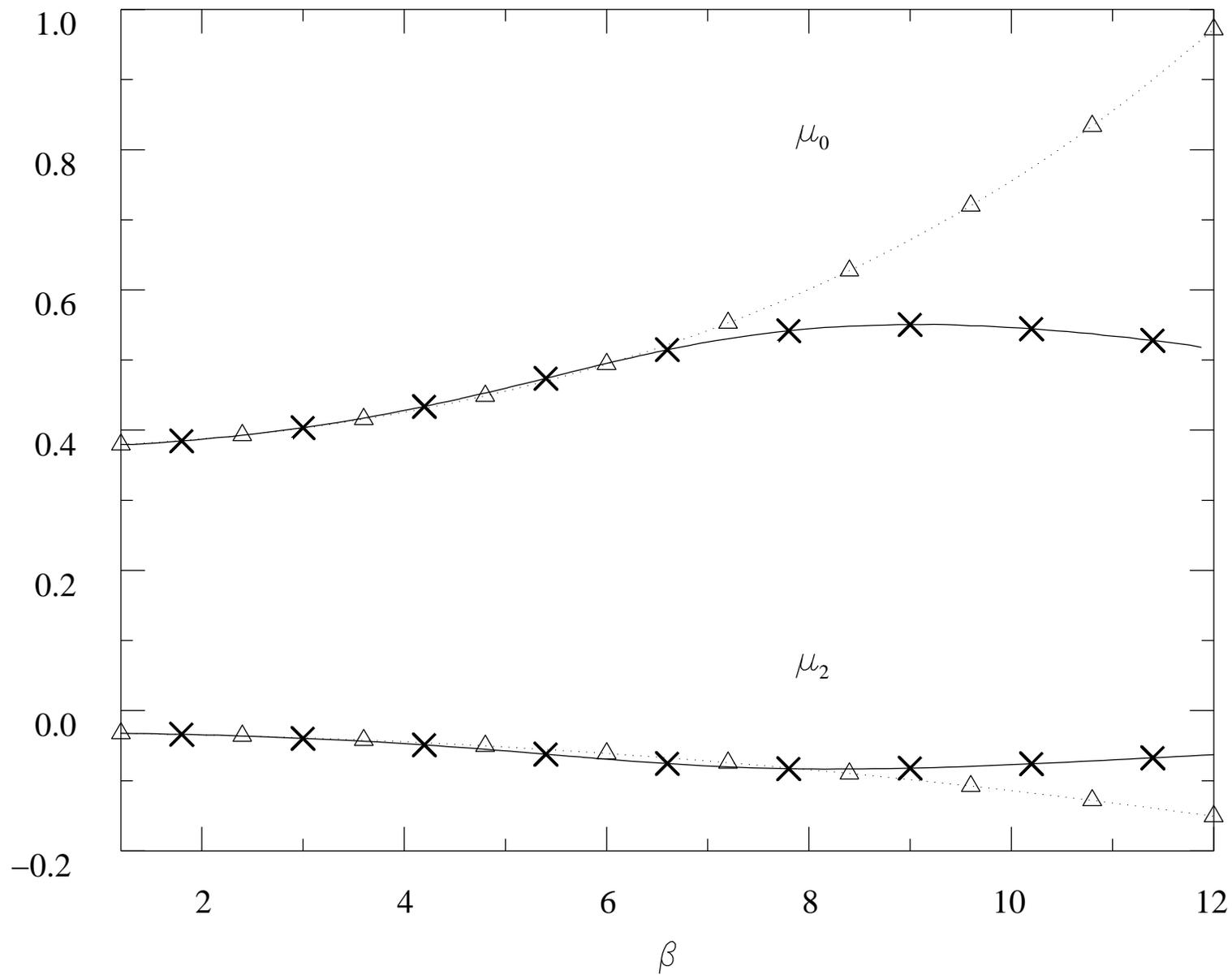

Fig. 3

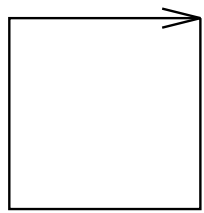

Fig. 2.1

$G_{2,1}$ 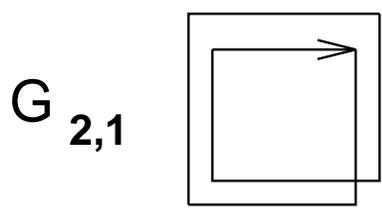

$G_{2,4}$ 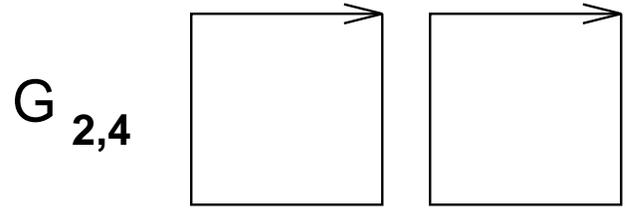

$G_{2,2}$ 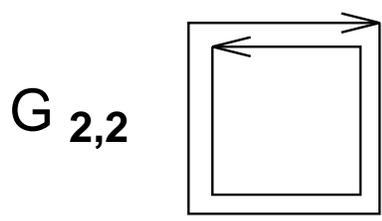

$G_{2,5}$ 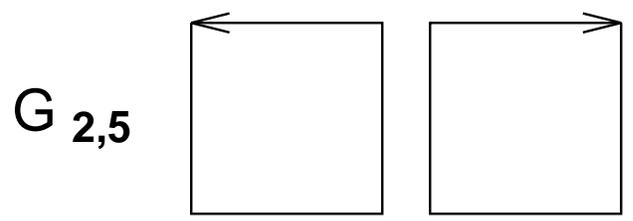

$G_{2,3}$ 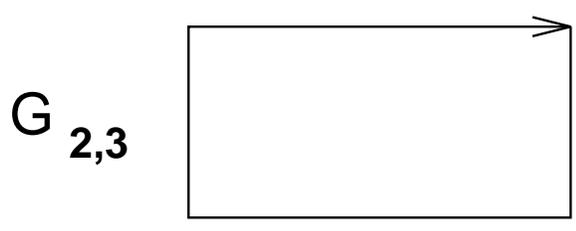

$G_{2,6}$ 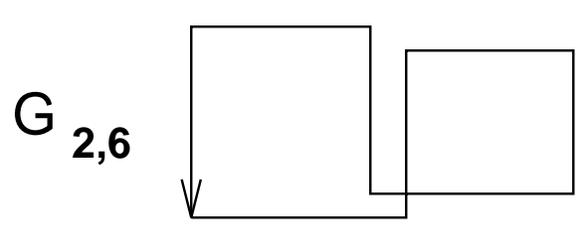

Fig. 2.2

G $_{3,1}$ 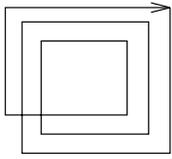

G $_{3,6}$ 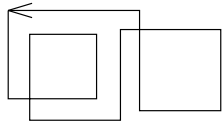

G $_{3,11}$ 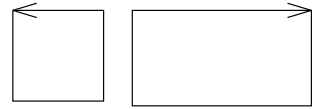

G $_{3,2}$ 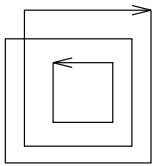

G $_{3,7}$ 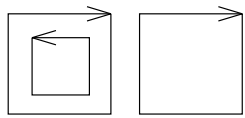

G $_{3,12}$ 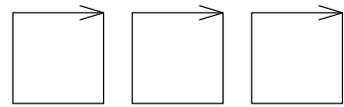

G $_{3,3}$ 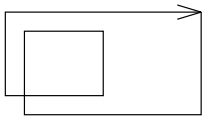

G $_{3,8}$ 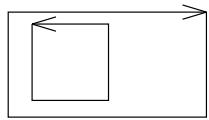

G $_{3,13}$ 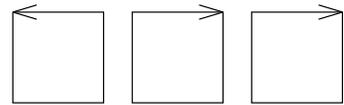

G $_{3,4}$ 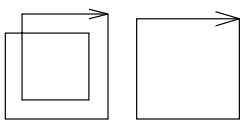

G $_{3,9}$ 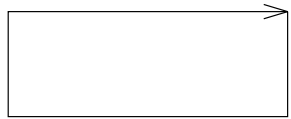

G $_{3,14}$ 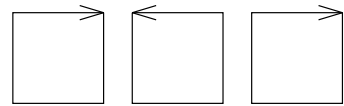

G $_{3,5}$ 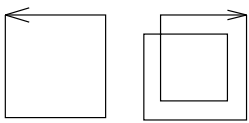

G $_{3,10}$ 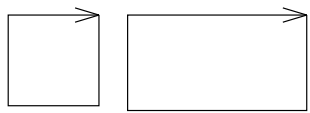

G $_{3,15}$ 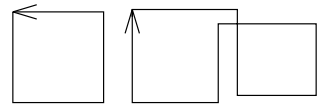

Fig. 2.3(a)

$G_{3,16}$

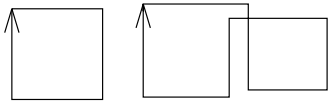

$G_{3,17}$

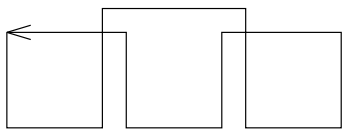

$G_{3,18}$

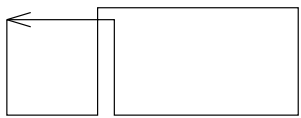

$G_{3,19}$

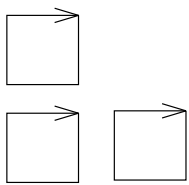

$G_{3,20}$

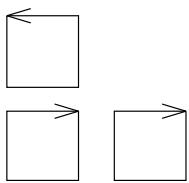

$G_{3,21}$

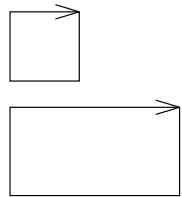

$G_{3,22}$

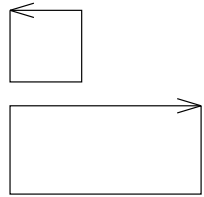

$G_{3,23}$

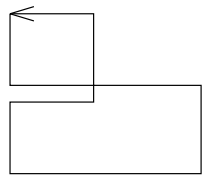

$G_{3,24}$

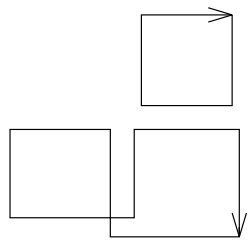

$G_{3,25}$

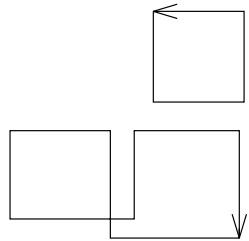

$G_{3,26}$

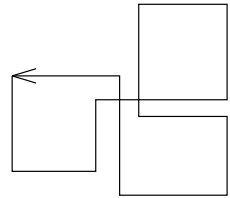

$G_{3,27}$

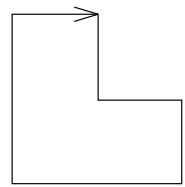

$G_{3,28}$

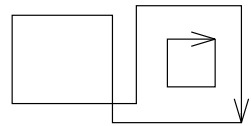

$G_{3,29}$

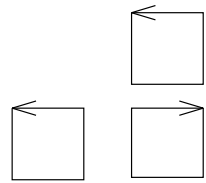

Fig. 2.3(b)